\documentclass[a4paper]{article}

\usepackage{multicol}
\usepackage{amsthm}
\usepackage{bm}

\usepackage[dvipdfmx]{graphicx}
\usepackage{bmpsize}
\usepackage{amssymb}
\usepackage{amsmath}
\usepackage{color}

\definecolor{dgreen}{rgb}{0.0, 0.5, 0.0}

\begin{document}
\fontsize{14pt}{28pt}\selectfont

\begin{center}
\bf{Periodicity of limit cycles 
in a max-plus dynamical system}
\end{center}\ \\
\fontsize{12pt}{12pt}\selectfont
\begin{center}
Yoshihiro Yamazaki and Shousuke Ohmori$^{*)}$\\ 
\ \\
\it{Department of Physics, Waseda University, Shinjuku, Tokyo 169-8555, Japan}\\
\ \\
*corresponding author: 42261timemachine@ruri.waseda.jp\\
\end{center}

\fontsize{11pt}{13pt}\selectfont\noindent

\baselineskip 30pt

{\bf Abstract}

By introducing a max-plus dynamical system 
having limit cycles, we discuss their periodicity, 
especially the number of discrete states in them.
We also find that quasi-periodic cycles exist  
depending on the bifurcation parameter in the system. 
Approximate relations between the number of states 
in the limit cycles 
and the value of the bifurcation parameter are proposed.\\
%
%\noindent
%{\bf Key Words} : ultradiscretization, bifurcation, normal forms, discrete dynamical system\\

\hrulefill

%%%%%%%%%%%%%%%%%%%%%%%%%%%%%%%%%%%%%%%%%%%%%%%%%%%%%%%%%%

%%%%%%%%%%%%%%%%%%%%%%%%%%%%%%%%%%%%%%%%%%%%%%%
%\section{Introduction}
%\label{sec:intro}

%\paragraph{background} \ 
% * max-plus は離散力学系に対して、時空間パターンを理解する上での新しい視点を与える可能性がある。
%
% * 力学系に基づく状態遷移
%
% * 偏微分方程式に基づく・・・。非線形波動や散逸構造。
%
% * pattern formation も
%
% * 一方、セルオートマトン。
%
% * この中間状態に位置するのがmax-plus
%
% * 差分方程式を出発点とすれば、一方の極限が微分であり、もう一方の極限が超離散、
% それから、更に状態の遷移ルールを抽出したセルオートマトンということになる。
%
% * max-plus も極限操作によって得られる関係式で、
% max演算を持つ区分線形な力学系として捉えることができる。 

For nonlinear and nonequilibrium phenomena, 
their description based on max-plus algebra has been made.
Soliton behaviors in integrable systems \cite{Tokihiro1996}, 
reaction-diffusion dynamics 
in dissipative systems \cite{Matsuya2015,Murata2013}, 
and bifurcation phenomena in dynamical systems 
\cite{Ohmori2020, Ohmori2021a} are typical examples.
Max-plus equations can be derived from discrete difference equations 
through ultradiscretization\cite{Tokihiro1996}.
They can be also obtained from continuous differential equations 
by appropriate discretization 
such as tropical discretization\cite{Murata2013}.
The crucial point is that there are cases 
where max-plus description can retain 
and elucidate essential dynamical structures 
of the original discrete or continuous systems.
%
%\cite{Gibo2015, Takahashi2001}

%-----------------
%\paragraph{Our recent works} \ 

%* 最近、我々は１次元で・・・、２次元で・・・を調べた。\\
%* Ohmori and Yamazaki \cite{Ohmori2021a}で提案した 
%ultradiscrete Hopf bifurcation を起こす max-plus equations,\\
%
Recently, we have derived the following max-plus 
dynamical system from the tropically discretized Sel'kov model 
via ultradiscretization\cite{Ohmori2021a}: 
\begin{equation}
	\begin{cases}
	\; X_{n+1} & = Y_n + \max(0,2X_n), \\
	\; Y_{n+1} & = B-\max(0,2X_n).
	\end{cases}
	\label{eqn:protoHopf}
\end{equation}
%
% * eq.(\ref{eqn:protoHopf})において、
% $B>0$のときリミットサイクル解が得られることを示した。
% 特に、その周期が7であることが示された。\\
We found that eq.(\ref{eqn:protoHopf}) has two limit cycles, 
$C$ and $C_{s}$, with period seven when $B>0$; 
they possess different basins.
%
% * 先行研究で、２種類のリミットサイクル$C$, $C_{s}$の存在することを示した。
%
% * $C$と$C_{s}$の違いは、region II-1 の点を保有するかどうかで、
% $C$は含み、$C_{s}$は含まない。\\
The difference between them is that 
$C$ has points in the region $X_{n} \leq 0$ and $Y_{n} \leq 0$, 
but $C_{s}$ does not.
%
%* この論文では、$C$についての解析を行う。\\
So far, understanding the periodicity of these limit cycles 
is not sufficient.
For example, it is not clear how the number 
of discrete states in the limit cycles is determined.
In this letter, we discuss such a unclear point 
by focusing on $C$.
%
% * この式において、$X_{n}/B \rightarrow X_{n}$、
% $Y_{n}/B \rightarrow Y_{n}$と変数変換しても、
% 系の本質的なdynamical behavior は変わらない。
In eq.(\ref{eqn:protoHopf}), 
we can perform the variable transformation, 
$X_{n}/B \rightarrow X_{n}$ and $Y_{n}/B \rightarrow Y_{n}$, 
without essential change of its dynamical properties 
for positive $B$.
%
% * つまり、eq.(\ref{eqn:protoHopf})において、
% リミットサイクルの挙動に着目するのであれば、
% 一般性を失うことなく、$B=1$とすることができる。
In other words, we can set $B=1$ in eq.(\ref{eqn:protoHopf}) 
without loss of generality if only $B > 0$ is treated.
%
%===============================
% \section{A generalization}
% \label{sec:generalization}
%
% * ここで、eq.(\ref{eqn:protoHopf})を一般化した
% 次の方程式を考える。\\
% * 本論文では、式(\ref{eqn:genHopf})による
% 動力学的挙動の$R$依存性について議論する。
Then we consider the following set of equations hereafter: 
\begin{equation}
	\begin{cases}
	\; X_{n+1} & = Y_n + \max(0,RX_n), \\
	\; Y_{n+1} & = 1-\max(0,RX_n).
	\end{cases}
	\label{eqn:genHopf}
\end{equation}
%
%* この式では新たなパラメータとして、$R>0$が導入されている。
Eq.(\ref{eqn:genHopf}) possesses a new parameter $R (>0)$ 
and is considered as a generalization of eq.(\ref{eqn:protoHopf}) 
with $B > 0$.
%
% \begin{itemize}
% 	\item リミットサイクルが起こる$R$の範囲を特定する。
% 	\item 周期が$R$に依存することがわかった。その依存性は？
% 	\item 周期性の特徴（２重周期の存在など・・・）
% 	\item 近似式の提案
% \end{itemize}

%%%%%%%%%%%%%%%%%%%%%%%%%%%%%%%%%%%%%%%%%%%%%%%
%\section{Dynamical properties of eq.(\ref{eqn:genHopf})}
%\label{sec:searchR}

\begin{figure}[h!]
	\begin{center}
	\includegraphics[width=10cm]{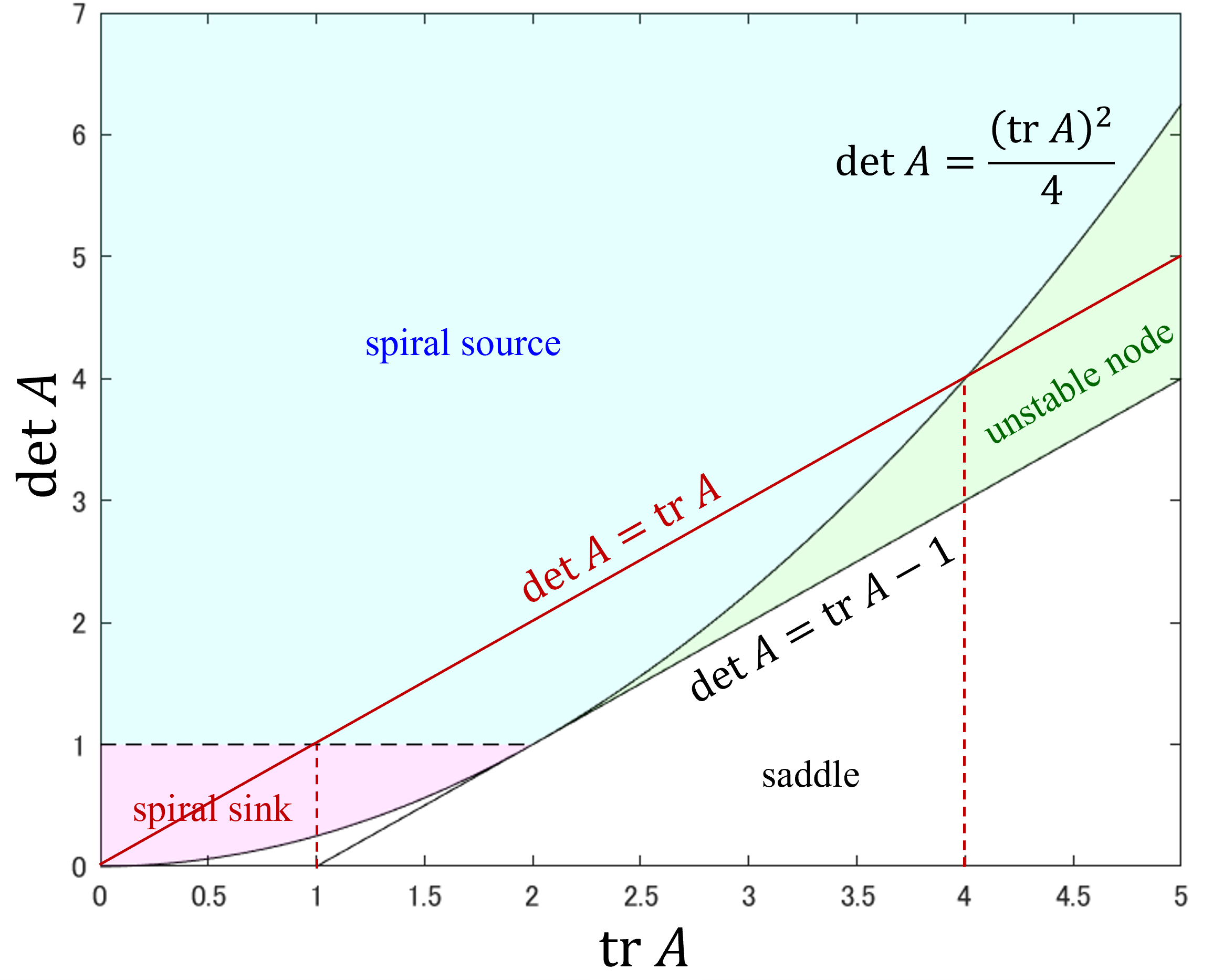}
	\caption{A diagram for dynamics 
		of the discrete linear dynamical system 
		$\bm{x}_{n+1} = A\bm{x}_{n}$ in two dimensions.}
	\label{fig:trAdetA}
	\end{center}
\end{figure}

\begin{figure}[h!]
	\begin{center}
	\includegraphics[width=6cm]{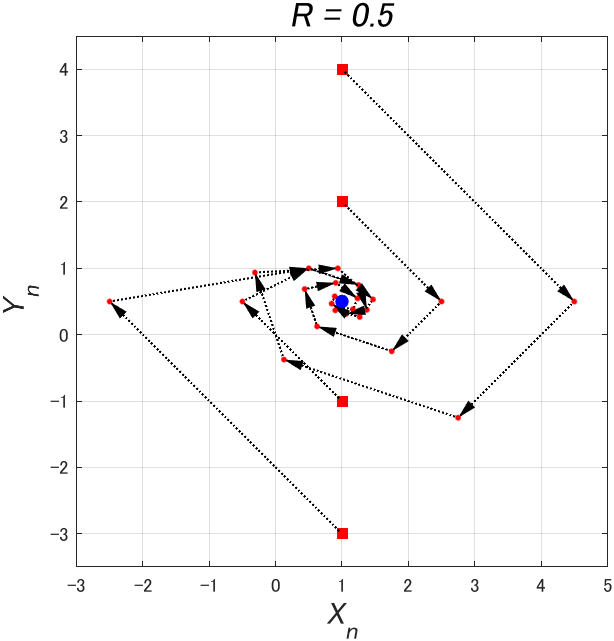}~~~~~
	\includegraphics[width=6cm]{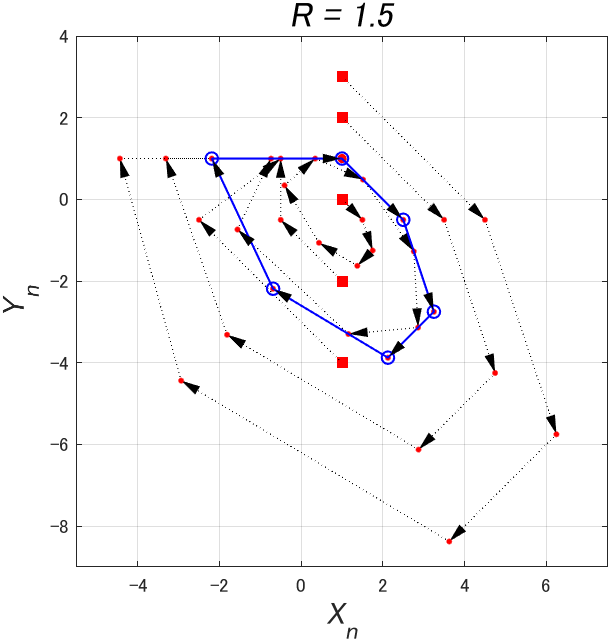}\\
	\hspace{0.7cm} (a) \hspace{7cm} (b)
	\caption{Examples of trajectories starting from red squares.
		(a) $R=0.5$. The blue point shows (1, 1/2) 
		which is the stable fixed point. 
		(b) $R=1.5$. The blue cycle shows the limit cycle $C$ 
		with period six.}
	\label{fig:trajectories}
	\end{center}
\end{figure}

% * 先行研究と同様、$(X_{n}, Y_{n})$ plane を
% Fig.\ref{fig:regions}のように3つの領域に分割し、それぞれ、
% region I, II-1, II-2 とする。\\
%According to the previous study\cite{Ohmori2021a}, 
Now we consider dynamical properties of eq.(\ref{eqn:genHopf}) 
by dividing $(X_{n}, Y_{n})$ plane 
into the three regions I ($X_{n} > 0$), 
II-1 ($X_{n} \leq 0, Y_{n} \leq 0$), 
and II-2 ($X_{n} \leq 0, Y_{n} > 0$).
%
% \begin{figure}[b!]
% 	\begin{center}
% 	\includegraphics[width=4cm]{regions.png}
% 	%
% 	\caption{The three regions in $(X_{n}, Y_{n})$ plane. 
% 		These regions are determined dependent on the signs 
% 		of $X_{n}$ and $Y_{n}$.}
% 	\label{fig:regions}
% 	\end{center}
% \end{figure}
%
% * region I においては、eqn.(\ref{eqn:genHopf})は
% 次のような行列表記をすることができる。
In region I, eq.(\ref{eqn:genHopf}) is represented 
by the matrix form, 
\begin{equation}
	\bm{x}_{n+1}
	=
	\left(
		\begin{array}{ccc}
			R & 1  \\
			-R & 0  \\
		\end{array}
	\right)
	\bm{x}_{n}
	+
	\left(
		\begin{array}{ccc}
			0  \\
			1  
		\end{array}
	\right), 
	\label{eqn:genHopf_1}
\end{equation} 
where 
$\bm{x}_{n} = \left(
	\begin{array}{c}
		X_{n} \\
		Y_{n} 
	\end{array}
\right)$. 
%
% * Eq.(\ref{eqn:genHopf_1})における解の挙動については、
% \cite{Galor}(p.90)に基づいて、
% その$R$依存性を次のように解析することができる。\\
$R$ dependence of dynamical properties of eq.(\ref{eqn:genHopf_1}) 
is understood as follows\cite{Galor}.
%
% * 先ず、Eq.(\ref{eqn:genHopf_1})の行列
% $
% 	A = \left(
% 		\begin{array}{ccc}
% 			R & 1  \\
% 			-R & 0  \\
% 		\end{array}
% 	\right)
% $
% に対して、その trace と determinant はそれぞれ、
% $\mathrm{tr}A = \mathrm{det}A = R$となる。
For the matrix 
$
	A = \left(
		\begin{array}{ccc}
			R & 1  \\
			-R & 0  \\
		\end{array}
	\right)
$, trace and determinant of $A$ are 
$\mathrm{tr}A = \mathrm{det}A = R$.
Figure \ref{fig:trAdetA} shows a typical diagram 
for the two dimensional dynamics $\bm{x}_{n+1} = A\bm{x}_{n}$ 
in terms of $\mathrm{tr}A$ and $\mathrm{det}A$.
%
%
% * このとき、動的挙動は$(\mathrm{tr}A, \mathrm{det}A)$plane上で
% Fig.\ref{fig:trAdetA}のように変化することがわかる。\\
It is found that the dynamical properties  
of eq.(\ref{eqn:genHopf_1}) depend 
on $R$ along the line $\mathrm{tr}A = \mathrm{det}A$. 
%
% * したがって、リミットサイクルが存在するのであれば、
% 挙動がspiral source となる$1 \leq R < 4$に存在することになる。\\
% * この場合、spiral は時計回りとなる。\\
From this figure, if limit cycles exist, 
$R$ is in the region $1 < R < 4$, 
where the fixed point of eq.(\ref{eqn:genHopf_1}) 
becomes clockwise spiral source.
%
% * region I の点（状態）はいずれ、region II-1, または、II-2 に入ることになるが、
% region II-2 に入った場合も再度region I に含まれ、
% いずれ、region II-1 に入ることになる。
Then there is a case for $1 < R < 4$ 
where a state (point) in region I finally gets into region II-1 
during time evolution of $\bm{x}_{n}$ 
by eq.(\ref{eqn:genHopf_1}).
%
% * region II-1 の任意の点$(X_0 \leq 0, Y_0 \leq 0)$はそれぞれ、
% $(X_0, Y_0) \rightarrow (Y_0, 1) \rightarrow (1, 1)$ となる。
Furthermore, any state in region II-1, 
$X_0 \leq 0$ and $Y_0 \leq 0$, 
changes as $(X_0, Y_0) \rightarrow (Y_0, 1) \rightarrow (1, 1)$ 
from eq.(\ref{eqn:genHopf}).
%
% * 以上より、$1 < R < 4$では
% $(1, 1)$を通るリミットサイクルが必ず存在することがわかる。
The state $(1, 1)$ is in region I, then 
there exists a cycle having the state $(1, 1)$ 
when $1 < R < 4$.
%
%* \txred{【注】$C_{s}$はregion II-1 を通らない。}
%
%* 以上の考察より、$R$がbifurcation parameter になっていることがわかる。//
Therefore, $R$ is considered as the bifurcation parameter 
for Neimark-Sacker bifurcation (Hopf bifurcation for 
discrete dynamical systems), which occurs at $R = 1$.
Figure \ref{fig:trajectories} shows trajectories 
for two different case of $R$:
(a) $R = 0.5$, (b) $R = 1.5$.
We can interpret this max-plus equation 
as describing a reset event from $(X_0, Y_0)$ to $(1, 1)$ 
when the value of $X_{n}$ becomes negative; 
the value `0' in the term $\max(0, RX_{n})$ is considered as 
the threshold for $X_{n}$. 

\clearpage
%%%%%%%%%%%%%%%%%%%%%%%%%%%%%%%%%%%%%%%%%%%%%%%
%\section{Periodicity of the limit cycles}
%\label{sec:periodicity}

%
% * region II-1 に入れば、2ステップで必ず$(1, 1)$になるので、
% リミットサイクルの周期はregion I に含まれる状態数を$n$とすると、
% $n+2$ で与えられることになる。
Since it takes two steps to reach the state $(1, 1)$ 
from any point in region II-1, 
%as shown in sec.\ref{sec:searchR}, 
the period of the limit cycle is $n+2$ 
where $n$ is the number of states in region I.
%
% * 実際、$R=2$ のとき、先行研究で示したように7周期であったので、
% $n=5$となっていることがわかる。
Actually, the previous study\cite{Ohmori2021a} shows 
$n=5$ for $R=2$, then the period of the limit cycle is 7.
%
% * そこで、$(1, 1)$を初期状態($\bm{x}_{0}$)として、
% eq.(\ref{eqn:genHopf_1}) で表される region I 中の時間発展$\bm{x}_{n}$
% を考える。
Here we consider time evolution of eq.(\ref{eqn:genHopf_1}) 
in region I starting from the initial state 
$\bm{x}_{0} = \left(
	\begin{array}{c}
		1 \\
		1 
	\end{array}
\right)$.
%
% * $\bm{x}_{0}$を出発点として、
% region I から region II-1 へ出るまでのステップ数$n$が求まれば、
% 周期は$n+2$で与えられることになる。
%
%* eq.(\ref{eqn:genHopf_1}) を形式的に解けば、
Eq.(\ref{eqn:genHopf_1}) is formally solved as 
\begin{equation}
	\bm{x}_{n} = A^{n}\bm{x}_{0} 
		+ \left(1 - A^{n}\right) \left(1 - A\right)^{-1}\bm{b}.
	\label{eqn:formal_xn}
\end{equation}
Denoting 
$\bm{x}_{n} = \left(
	\begin{array}{c}
		G_{X}(n, R) \\
		G_{Y}(n, R)
	\end{array}
\right)$
%$\displaystyle \bm{x}_{n} = \left( G_{X}(n, R), G_{Y}(n, R) \right) $
as the solution of eq.(\ref{eqn:formal_xn}) 
by setting 
$
	A = \left(
		\begin{array}{ccc}
			R & 1  \\
			-R & 0  \\
		\end{array}
	\right)
$, 
$\displaystyle \bm{b} = \left(
	\begin{array}{c}
		0 \\
		1 
	\end{array}
\right)$, and 
$\bm{x}_{0} = \left(
	\begin{array}{c}
		1 \\
		1 
	\end{array}
\right)$, 
$G_{X}(n, R)$ and $G_{Y}(n, R)$ are explicitly given as 
%
%* ここで、$G_{X}(n, R)$、$G_{Y}(n, R)$はそれぞれ、以下の通り。
% \begin{eqnarray}
% 	G_{X}(n, R) & = & 1 
% 		- \frac{2^{-n}\sqrt{R} \left( -\sqrt{-4R+R^{2}} +R \right)^{n}}{\sqrt{-4+R}}
% 		+ \frac{2^{-n}\sqrt{R} \left(  \sqrt{-4R+R^{2}} +R \right)^{n}}{\sqrt{-4+R}}
% 	\label{eqn:Gx_nR}
% \end{eqnarray}
\begin{eqnarray}
	G_{X}(n, R) & = & 1 
		+ \frac{R^{\frac{n+1}{2}}}{2^{n}i\sqrt{4-R}}
			\biggl\{ \left( \sqrt{R} + i \sqrt{4-R} \right)^{n}
			-\left( \sqrt{R} - i \sqrt{4-R} \right)^{n} \biggr\}, 
	\label{eqn:Gx_nR}
\end{eqnarray}

% \begin{eqnarray}
% 	G_{Y}(n, R) & = & 1 - R
% 		- \frac{2^{-n}\sqrt{R} \left( -\sqrt{-4R+R^{2}} +R \right)^{n}}{\sqrt{-4+R}}
% 		+ \frac{2^{-n}\sqrt{R} \left(  \sqrt{-4R+R^{2}} +R \right)^{n}}{\sqrt{-4+R}}
% 		\\ \nonumber
% 	& & 
% 		+ 2^{-n-1} R \left( -\sqrt{-4R+R^{2}} +R \right)^{n}
% 		+ 2^{-n-1} R \left(  \sqrt{-4R+R^{2}} +R \right)^{n}
% 		\\ \nonumber
% 	& & 
% 		+ 2^{-n-2} \sqrt{-4R+R^{2}} \left( -\sqrt{-4R+R^{2}} +R \right)^{n}
% 		- 2^{-n-2} \sqrt{-4R+R^{2}} \left(  \sqrt{-4R+R^{2}} +R \right)^{n}
% 		\\ \nonumber
% 	& &
% 		+ \frac{2^{-n-2} R^{\frac{3}{2}} \left( -\sqrt{-4R+R^{2}} +R \right)^{n}}{\sqrt{-4+R}}
% 		- \frac{2^{-n-2} R^{\frac{3}{2}} \left(  \sqrt{-4R+R^{2}} +R \right)^{n}}{\sqrt{-4+R}}
% 	\label{eqn:Gy_nR}
% \end{eqnarray}
\begin{eqnarray}
	G_{Y}(n, R) & = & 1 - R
		+ \frac{R^{\frac{n+1}{2}}}{2^{n+1}}
		\Biggl[ \biggl\{ \left( \sqrt{R} + i \sqrt{4-R} \right)^{n}
		+\left( \sqrt{R} - i \sqrt{4-R} \right)^{n} \biggr\} 
	 \nonumber \\
	& & -\frac{1}{i}\sqrt{\frac{R}{4-R}} 
		\biggl\{ \left( \sqrt{R} + i \sqrt{4-R} \right)^{n}
		-\left( \sqrt{R} - i \sqrt{4-R} \right)^{n} \biggr\} \Biggr], 
	\label{eqn:Gy_nR}
\end{eqnarray}
$(1 < R < 4)$.
%* なお、$G_{X}$と$G_{Y}$の間には、次のような関係がある。
We note that the following relation holds 
between $G_{X}$ and $G_{Y}$, 
\begin{equation}
	R G_{X}(n-1, R) + G_{Y}(n, R) = 1.
	\label{eqn:GxGy_R}
\end{equation}

\begin{figure}[h!]
	\begin{center}
	\includegraphics[width=7.5cm]{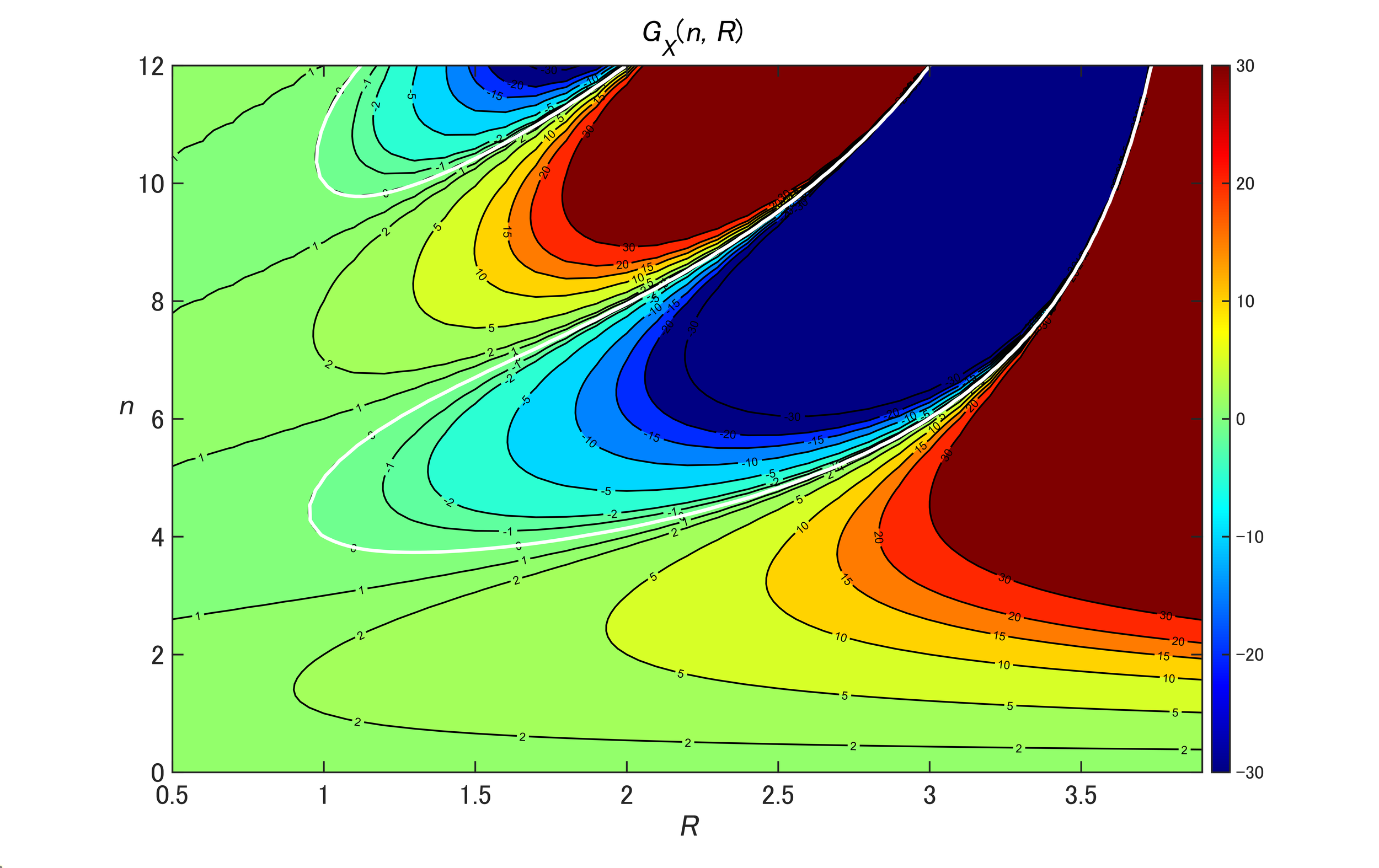}
	\includegraphics[width=7.5cm]{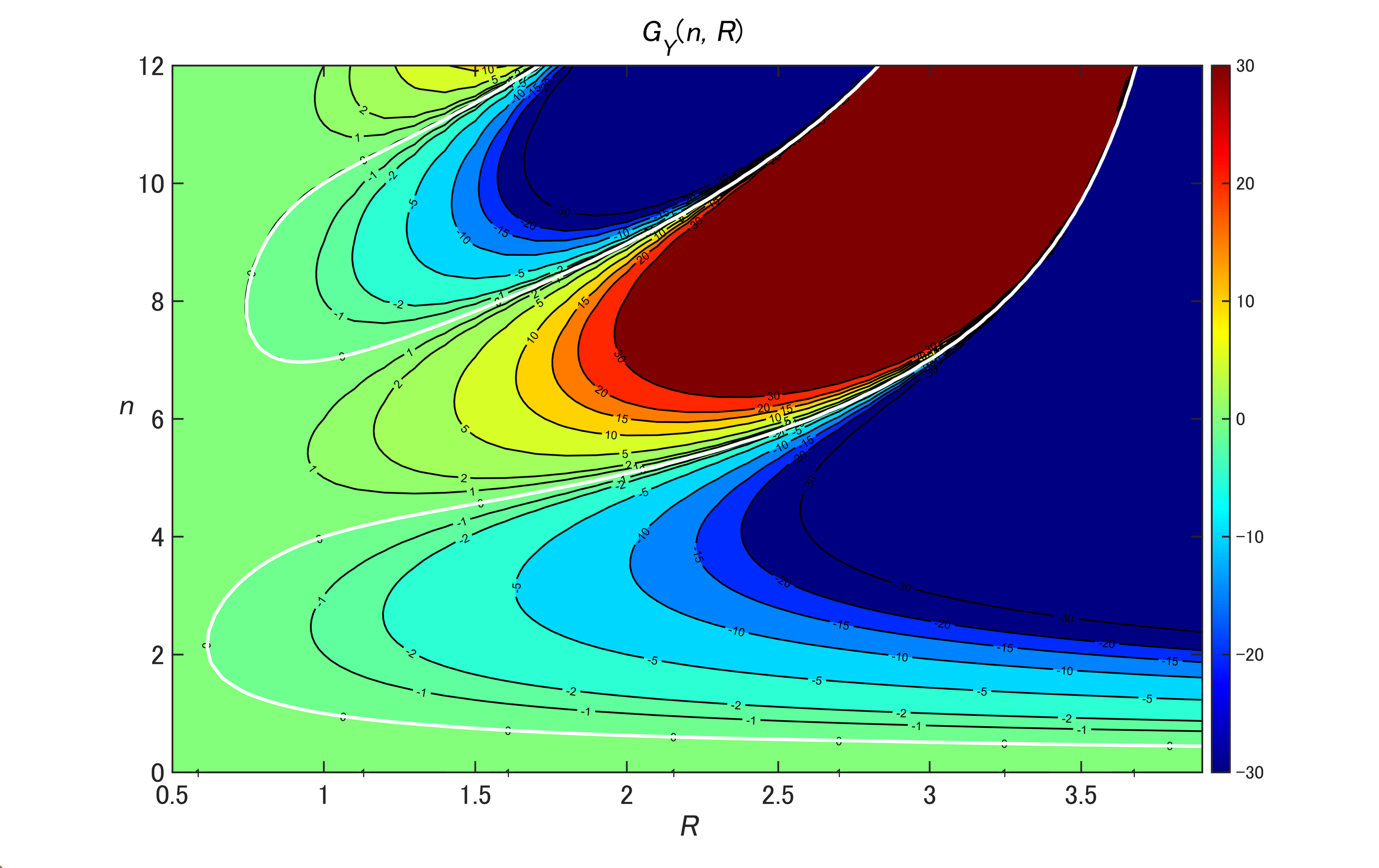}\\
	(a) \hspace{7.2cm} (b)
	\caption{Contour plots of (a) $G_{X}(n, R)$ 
		and (b) $G_{Y}(n, R)$. 
		The white curves show the relations of 
		(a) $G_{X}(n, R)=0$ and (b) $G_{Y}(n, R)=0$.}
	\label{fig:Gxy_nR}
	\end{center}
\end{figure}

\begin{figure}[h!]
	\begin{center}
	\includegraphics[width=8cm]{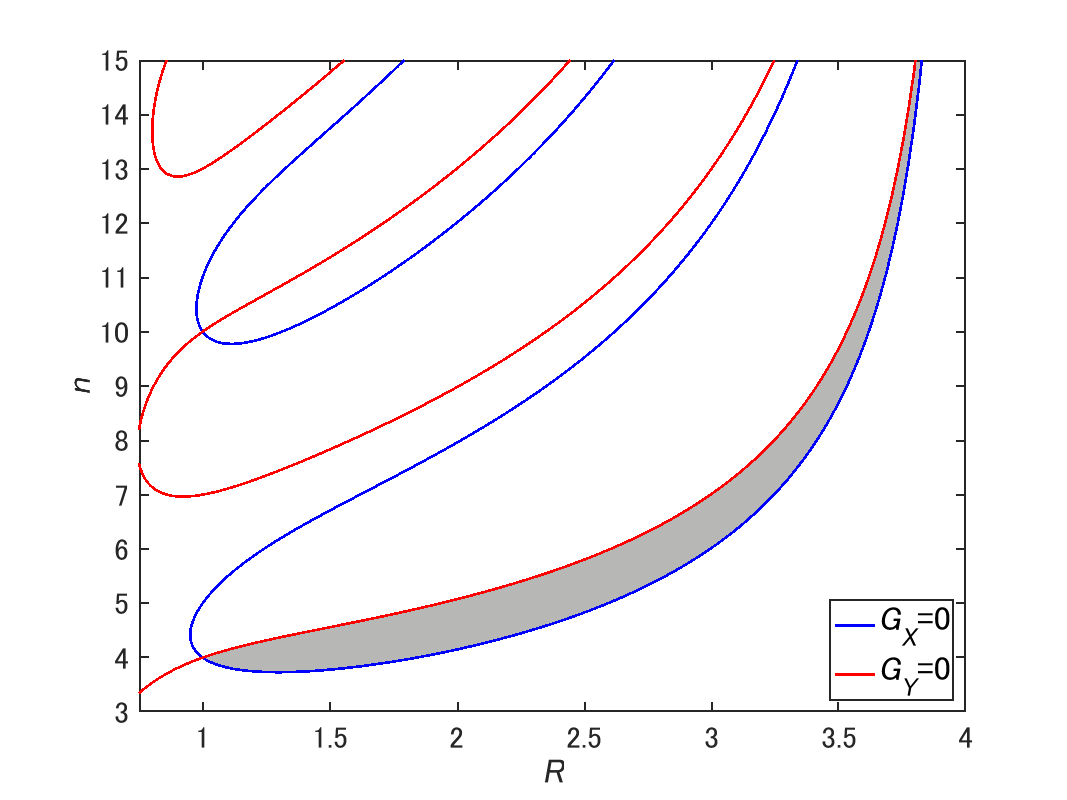}
	\caption{The curves for $G_{X}(n, R)=0$ and $G_{Y}(n, R)=0$. 
		The gray region shows the condition for existence 
		of limit cycles.}
	\label{fig:Rn_nulls}
	\end{center}
\end{figure}

%
% * したがって、$G_{X}(n, R) \leq 0$, $G_{Y}(n, R) \leq 0$となる
% 最小の$n$を求めれば良いことになる。
Based on eqs.(\ref{eqn:Gx_nR}) and (\ref{eqn:Gy_nR}), 
the number of states in region I of the limit cycle 
is given by the minimum $n$ satisfying 
$G_{X}(n, R) \leq 0$ and $G_{Y}(n, R) \leq 0$ for a fixed $R$, 
which means $\bm{x}_{n}$ is in region II-1.
%
% * ここで、$G_{X}(n, R)$、および、$G_{Y}(n, R)$のcontour plots
% はFig.\ref{fig:Gxy_nR}のようになる。
Figure \ref{fig:Gxy_nR} shows the contour plots 
of $G_{X}(n, R)$ and $G_{Y}(n, R)$.
%
% * また、$G_{X}(n, R)=0$、$G_{Y}(n, R)=0$となる
% 曲線がFig.\ref{fig:Rn_nulls}に示されている。
Figure \ref{fig:Rn_nulls} shows the curves 
for $G_{X}(n, R)=0$ and $G_{Y}(n, R)=0$, 
which are depicted as white curves in Fig.\ref{fig:Gxy_nR}.
%
% * Fig.\ref{fig:Gxy_nR}のcontour plots を考慮すると、
% この gray region に入る$(n, R)$において limit cycle が
% 実現されると考えられる。
It is found from the signs of $G_{X}$ and $G_{Y}$ 
shown in Fig.\ref{fig:Gxy_nR} 
that the limit cycles can emerge with $(n, R)$ 
in the gray region of Fig.\ref{fig:Rn_nulls}.
%
% * 実際に、ある与えられた$n$を満たす$R$の範囲を数えるということで進めると、
% $G_{X}(n, R)=0$、$G_{Y}(n, R)=0$を$R$について解いたときの解を
% それぞれ$R_{X}(n)$, $R_{Y}(n)$とすると、周期$n+2$となる$R$の範囲は、
% $R_{Y}(n) < R < R_{X}(n)$で与えられることになる。
Now we determine the region of $R$ 
satisfying $G_{X}(n, R) \leq 0$ and $G_{Y}(n, R) \leq 0$
for a given $n$.
When we express the solution of $G_{X}(n, R)=0$ and $G_{Y}(n, R)=0$ 
with respect to $R$ as $R_{X}(n)$ and $R_{Y}(n)$, 
the region of $R$ for existence of the limit cycle 
with period $n+2$ is $R_{Y}(n) \leq R \leq R_{X}(n)$.
%
%* 実際、表のように範囲を数値的に定めることができる。
%* limit cycle の周期は$n+2$で与えられたので、
%* まず、$R=1$、$n=4$のとき、$G_{X}=G_{Y}=0$
%* 最小の$n$
Table \ref{tbl:periodicity} shows the numerical results 
of such regions as a function of $n$.
\begin{table}[h!]
	\begin{center}
		\caption{Numerically obtained regions of $R$ 
			for existence of limit cycles with period $n+2$, 
			where $n$ is the number of states 
			in region I of the cycles.}
		\begin{tabular}{c|c|l}
	    $n$ & period($=n+2$) & $R_{Y}(n) \leq R \leq R_{X}(n)$ \\
			\hline
  	   4 &  6 & 1\color{white}.000000$\cdots $\color{black} $ \sim$ 1.83928$\cdots$ \\
  	   5 &  7 & 1.93318 $\cdots \sim$ 2.59205 $\cdots$\\
  	   6 &  8 & 2.60229 $\cdots \sim$ 2.99375 $\cdots$ \\
  	   7 &  9 & 2.99585 $\cdots \sim$ 3.24522 $\cdots$ \\
  	   8 & 10 & 3.24576 $\cdots \sim$ 3.41367 $\cdots$ \\
  	   9 & 11 & 3.41383 $\cdots \sim$ 3.53191 $\cdots$ \\
  	  10 & 12 & 3.53196 $\cdots \sim$ 3.61797 $\cdots$ \\
  	  \vdots & \vdots & \vdots
    \end{tabular}
		\label{tbl:periodicity}
	\end{center}
\end{table}

%\clearpage

%%%%%%%%%%%%%%%%%%%%%%%%%%%%%%%%%%%%%%%%%%%%%%%
%\section{Existence of quasi-periodic solutions}
%\label{sec:property}

\begin{figure}[b!]
	\begin{center}
		\includegraphics[width=6cm]{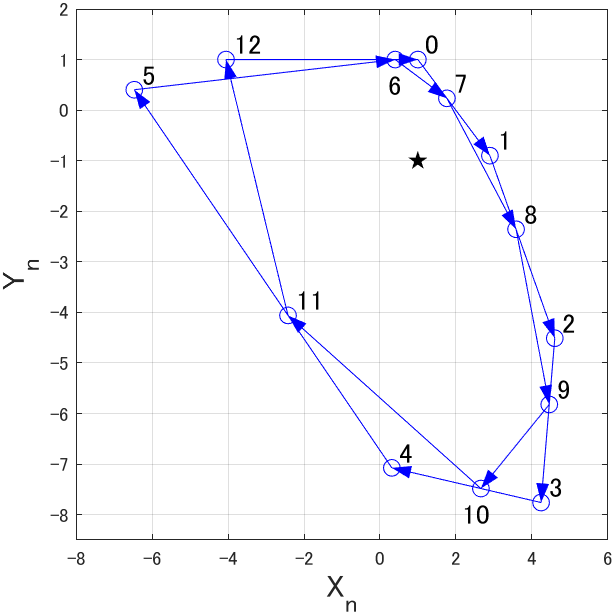}
		\includegraphics[width=8cm]{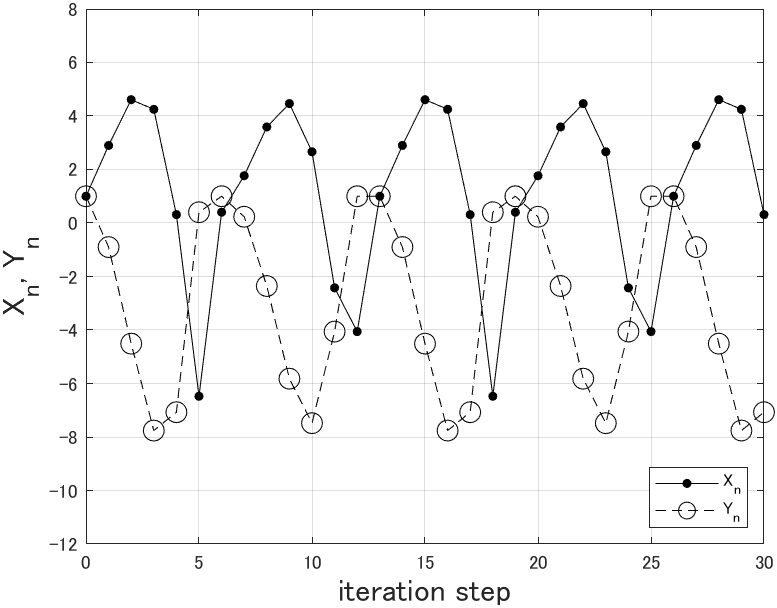}\\
		(a) \hspace{7cm} (b)\\
		\includegraphics[width=6cm]{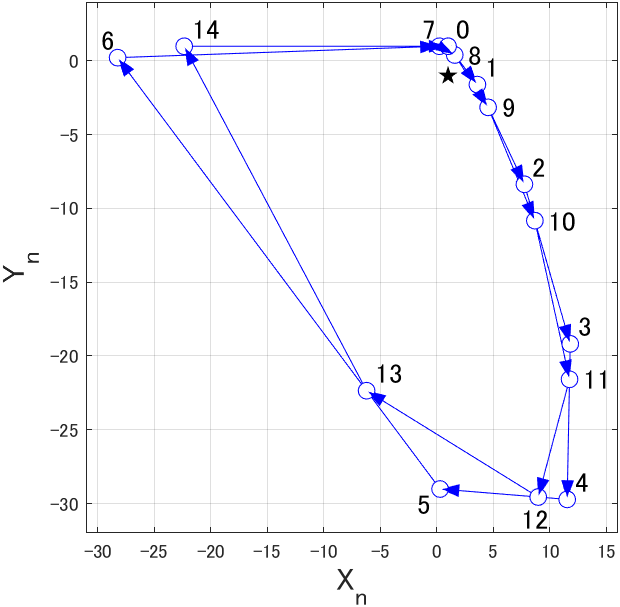}
		\includegraphics[width=8cm]{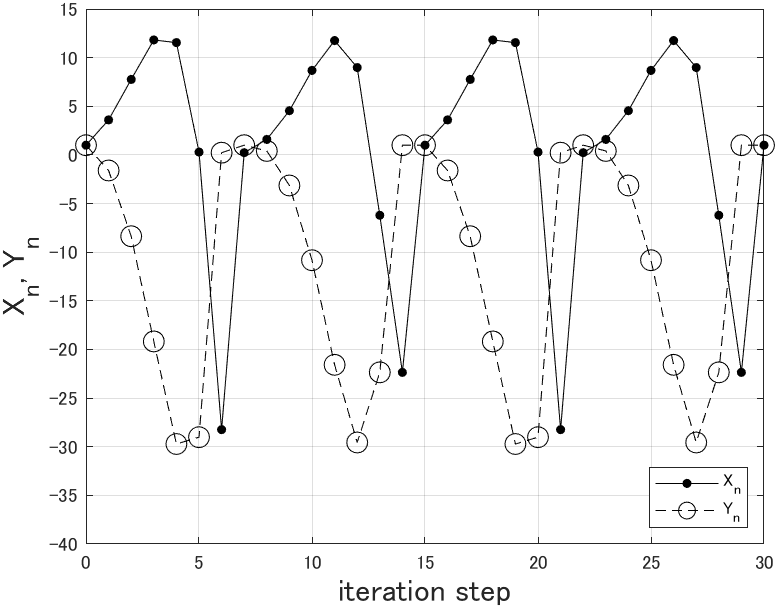}\\
		(c) \hspace{7cm} (d)\\
	\caption{Quasi-periodic cycles for two different values of $R$. 
		(a) trajectory and (b) time evolution for $R=1.9$.
		(c) trajectory and (d) time evolution for $R=2.6$.}
		\label{fig:qp_67}
	\end{center}
\end{figure}

%
% * ここで、Tbl.\ref{tbl:periodicity}をみると、
% それぞれの周期が出現する$R$の範囲の間に有限の間隔があることに気づくであろう。
Table \ref{tbl:periodicity} shows that 
there are finite gaps between regions of $R$ 
for two limit cycles with periods $n$ and $n+1$.
%
% * 例えば、周期6と周期7のcycleが出現する$R$の範囲の間に、
% 1.83928$\cdots \sim$ 1.93318 $\cdots$の領域が存在する。
For example, the region $1.83928 \cdots < R < 1.93318 \cdots$ 
corresponds to the gap between the two cycles 
with periods 6 and 7.
%
%* 我々は、この隙間の領域で準周期解が存在することを発見した。
In these gaps, $R_{X}(n) < R < R_{Y}(n+1)$, 
we find quasi-periodic limit cycles 
composed of $n$ and $n+1$ periods.
Figure \ref{fig:qp_67} shows the examples of 
the quasi-periodic cycles with $(6+7)$ period for $R = 1.9$ 
and with $(7+8)$ period for $R = 2.6$.
%
% * 証明はしていないが、周期$n$と$n+1$($n \geq 4$の自然数)の間で、
% $n+(n+1)$の準周期が存在すると思われる。
% * この隙間では、対応する自然数$n$が存在しないことを反映しているようだ。
Existence of such quasi-periodic limit cycles seems to be 
due to absence of the integer $n$ in these gaps.
%
%* なお、このような隙間が必ず存在することがわかる。
The gaps exist between every two limit cycles 
with periods $n$ and $n+1$.
%
% * 実際、$G_{X}(n, R_{X}(n))=0$ を満たす$R_{X}(n)$について、
% eq.(\ref{eqn:Gx_Gy_R})より、
% \[
% 	R G_{X}(n, R_{X}(n)) + G_{Y}(n+1, R_{X}(n)) = 
% 	G_{Y}(n+1, R_{X}(n)) = 1
% \]
% が成り立つ。
In proof, for $R=R_{X}(n)$ which satisfies $G_{X}(n, R_{X}(n))=0$, 
the following relation holds from eq.(\ref{eqn:GxGy_R}), 
\[
	R_{X}(n) G_{X}(n, R_{X}(n)) + G_{Y}(n+1, R_{X}(n)) = 
	G_{Y}(n+1, R_{X}(n)) = 1. 
\]
Then the relation 
\[
	G_{Y}(n+1, R_{X}(n)) = 1 > 0 = G_{Y}(n+1, R_{Y}(n+1)) 
\]
is obtained. 
%
% であり、Fig.\ref{fig:Gxy_nR}を考慮すると、
% \[
% 	R_{X}(n) < R_{Y}(n+1)
% \]
% の成り立つことが示される。
Considering the contour plots shown in Fig.\ref{fig:Gxy_nR}(b), 
we find
\[
	R_{X}(n) < R_{Y}(n+1), 
\]
which shows existence of the finite gaps 
for occurrence of quasi-periodic limit cycles.
%
% Further characterization of the quasi-periodic solutions 
% will be a future work.

%%%%%%%%%%%%%%%%%%%%%%%%%%%%%%%%%%%%%%%%%%%%%%%
%\section{Approximations for periodicity of the cycles}
%\label{sec:approximation}

\begin{figure}[h!]
	\begin{center}
	\includegraphics[width=8cm]{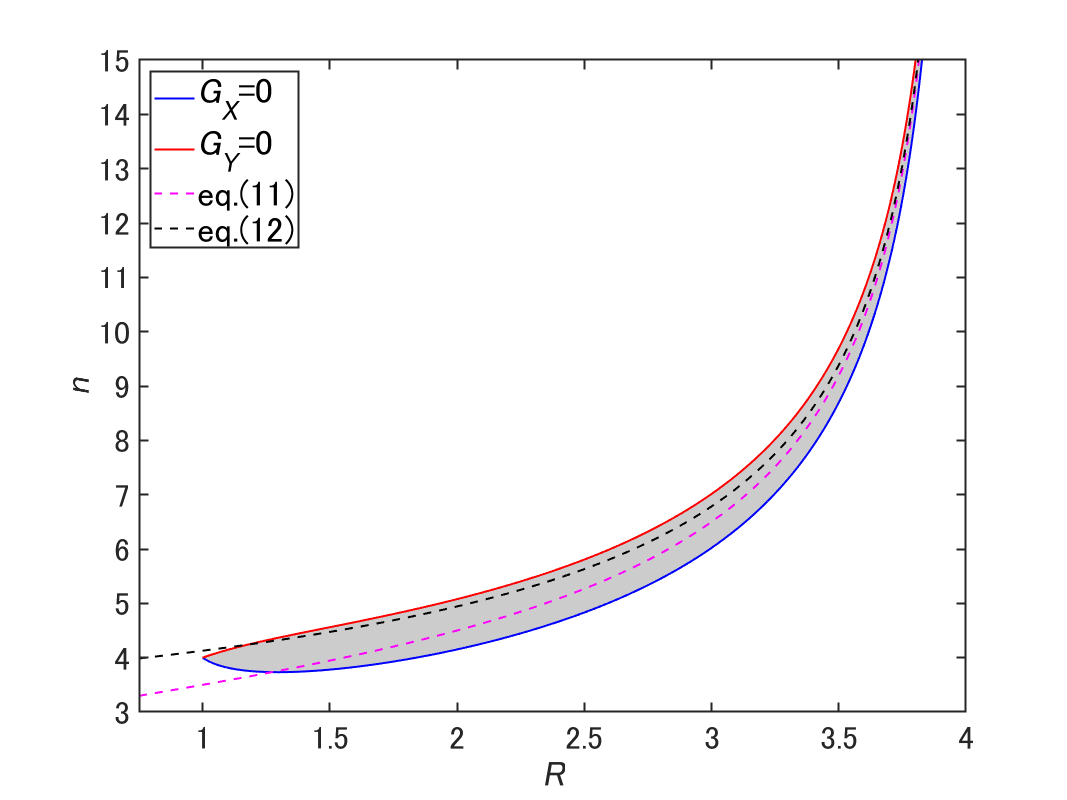}
	\caption{The two approximations given 
		by eqs.(\ref{eqn:approx}) and (\ref{eqn:approx1}).
		The gray region is the same as Fig.\ref{fig:Rn_nulls}.}
	\label{fig:Rn_nulls_a}
	\end{center}
\end{figure}

%
% * $n$ と$R$の関係は上記の通りだが、近似的な関係があると良いと思われる。
% * 範囲が広がっているので、近似式もいくつかの式が提案できる。
We propose several approximate relations 
between $n$ and $R$. 
%for occurrence of the cycles.
%
% * $\theta$表現は近似式を求めるときに役に立ちそうなので、
% ここに移動する。
In order to obtain them, 
the following variable transformation from $R$ to 
$\theta$ is considered, 
\begin{equation}
	e^{i\theta} = \frac{\sqrt{R} + i \sqrt{4-R}}{2}. 
	\label{eqn:R_theta}
\end{equation}
%
% * この変数変換において、$\theta$の取りうる範囲は
% $\displaystyle 0 < \theta < \frac{\pi}{3}$となる。
The range of values which $\theta$ can take is 
$\displaystyle 0 < \theta < \frac{\pi}{3}$ 
because of $1 < R < 4$.
%
% * 実際、$G_{X}(n, R)$、$G_{Y}(n, R)$に対して、
% 変数変換$R \rightarrow \theta$を行って得られる
% $G_{X}(n, \theta)$、$G_{Y}(n, \theta)$は、
Applying this variable transformation 
to eqs.(\ref{eqn:Gx_nR}) and (\ref{eqn:Gy_nR}), 
we obtain $G_{X}$ and $G_{Y}$ as a function of $n$ and $\theta$.
\begin{eqnarray}
	G_{X}(n, \theta) & = & 
		1 + \frac{ \left( 2 \cos \theta \right)^{n+1}}{\sin \theta}
		\sin n\theta, 
	\label{eqn:Gx_ntheta} \\
	G_{Y}(n, \theta) & = & 
		1 - 4 \cos^{2} \theta 
		+ \frac{ \left( 2 \cos \theta \right)^{n+2}}{\sin \theta}
		\sin (n-1) \theta, 
	\label{eqn:Gy_ntheta}
\end{eqnarray}
($n \geq 4$).
%
% * 式\ref{eqn:Gx_ntheta} より、
% Fig.\ref{fig:Gxy_nR} から、$G_{X}(n, \theta)=0$, 
% $G_{Y}(n, \theta)=0$ から離れると、急激に
% $|G_{X}|$, $|G_{Y}|$の値が大きくなることが示されている。
Figure \ref{fig:Gxy_nR} tells that 
the values of both $|G_{X}|$ and $|G_{Y}|$ rapidly increase 
when they move away from the curves $G_{X}=0$ and $G_{Y}=0$ 
especially for large $R$ (small $\theta$).
%
% * したがって、$G_{X}(n, \theta)=0$を満たす$\theta$
% を$G_{X}(n, \theta)=1$を満たす$\theta$で近似すると、
% これは、$\sin n\theta = 0$を満たす。
Therefore, the value of $\theta$ for $G_{X}(n, \theta)=0$ 
can be approximated by $\theta_{X}$ 
satisfying $G_{X}(n, \theta_{X})=1$, namely 
$\sin n\theta_{X} = 0$ from eq.(\ref{eqn:Gx_ntheta}).
%
% * $n=0$ のとき$G_{X}(n, \theta)=1$なので、
% 次は、$n\theta=\pi$のときである。
As the smallest positive $n$ for $\sin n\theta_{X} = 0$, 
we obtain $n\theta_{X}=\pi$.
%
% * 一方、同様に、$G_{Y}(n, \theta)=0$を満たす$\theta$
% を$G_{X}(n, \theta)=1-4\cos^{2}\theta$を満たす$\theta$で近似すると、
% これは、$\sin (n-1)\theta = 0$を満たす。
%* したがって、$(n-1)\theta = \pi$となる。
%
In the similar way for eq.(\ref{eqn:Gy_ntheta}), 
we also obtain the approximate value $\theta_{Y}$ 
which satisfies $\sin (n-1)\theta_{Y} = 0$, 
that is $(n-1)\theta_{Y} = \pi$.
Considering the forms of $\theta_{X}$ and $\theta_{Y}$, 
we roughly estimate $\theta$ as 
$\displaystyle \left( n-\frac{1}{2} \right) \theta \approx \pi$.
(The constant $\displaystyle \frac{1}{2}$ is just a candidate 
between 0 and 1. 
A better constant may be found from an appropriate fitting.)
Then an approximate relation between $n$ and $\theta$ (or $R$)
is obtained as  
%
%* この２つの条件の中点を採用して、$n$と$\theta$(したがって、$R$)の関係は
\begin{equation}
	n \approx \frac{1}{2} + \frac{\pi}{\theta}
	= \frac{1}{2} + \frac{\pi}{\arccos \frac{\sqrt{R}}{2}}.
	\label{eqn:approx}
\end{equation}
%と近似的に考えることができる。
%
From the relation $\displaystyle \cos^{2}\theta = \frac{R}{4}$, 
$\theta$ is further approximated as 
$\displaystyle \theta \approx \frac{\sqrt{4-R}}{2}$ 
for $\theta \ll 1$ ($R \lesssim 4$). 
Therefore, eq.(\ref{eqn:approx}) can be rewritten as 
%
% * また、$\displaystyle \cos^{2}\theta = \frac{R}{4}$であり、
% $\theta \ll 1$では、
% $\displaystyle \theta \approx \frac{\sqrt{4-R}}{2}$と近似できるので、
% eqn.(\ref{eqn:approx})は、
\begin{equation}
	n \approx \frac{1}{2} + \frac{2\pi}{\sqrt{4-R}}. 
	\label{eqn:approx1}
\end{equation}
% とも近似できそうである。
%when $\theta \ll 1$.
%
%* Fig.\ref{fig:Rn_nulls_a}中の点線が、これらの近似式を表している。
%
Figure \ref{fig:Rn_nulls_a} shows the plots of 
eqs.(\ref{eqn:approx}) and (\ref{eqn:approx1}) 
together with the gray region in Fig.\ref{fig:Rn_nulls}.
%
% * どちらの近似式も領域内には概ね収まっており、特に$R$が4に近づくにつれて、
% 良い一致を示している。
This figure shows that these two approximate relations 
are roughly in the gray region; 
these approximations are found to be well 
especially with larger $R$.

%%%%%%%%%%%%%%%%%%%%%%%%%%%%%%%%%%%%%%%%%%%%%%%
%\section{Conclusion}
%\label{sec:conclusion}

In conclusion, we have discussed periodicity 
of the limit cycles based on eq.(\ref{eqn:genHopf}).
This equation has the bifurcation parameter $R$, 
and the limit cycles emerge when $1 < R < 4$.
The number of states $n$ in the limit cycles is given 
as a function of $R$. 
It is found that quasi-periodic cycles exist 
depending on the value of $R$.
The two approximations for the relation 
between $n$ and $R$, i.e. 
eqs.(\ref{eqn:approx}) and (\ref{eqn:approx1}), 
have been demonstrated.
The present results are expected to give 
fundamental information for periodic phenomena 
with max-plus description. 

\bigskip

\noindent
{\bf Acknowledgement}

The authors are grateful to 
% Assoc. prof. M. Murata, at Tokyo University of Agriculture and Technology, 
% Assoc. prof. K. Matsuya at Musashino University,
% Prof. D. Takahashi, 
Prof. T. Yamamoto, and Prof. Emeritus A. Kitada 
at Waseda University for encouragements. 
%
%Also, we greatly appreciate the valuable comments from an reviewer.
This work was
supported by Sumitomo Foundation, Grant Number 200146.

\bigskip

%\noindent
%{\bf Data Availability}

%Data sharing is not applicable to this article as no new data were created or analyzed in this study.\\


\begin{thebibliography}{9}

\bibitem{Tokihiro1996} %soliton
T. Tokihiro, D. Takahashi, J. Matsukidaira, and J. Satsuma, Phys. Rev. Lett. {\bf 76}, 3247 (1996).

\bibitem{Matsuya2015} %RD
K. Matsuya and M. Murata, Discrete Contin. Dyn. Syst. B {\bf 20} 173 (2015).

\bibitem{Murata2013}
M. Murata, J. Differ. Equations Appl. {\bf 19} 1008 (2013).

\bibitem{Ohmori2020}
S. Ohmori and Y. Yamazaki, J. Math. Phys. {\bf61} 122702 (2020)

\bibitem{Ohmori2021a}
S. Ohmori and Y. Yamazaki, submitted (arXiv:2107.02435).

\bibitem{Galor}
O. Galor, \emph{Discrete Dynamical Systems} (Springer, New York 2010).



% \bibitem{Gibo2015}
% S. Gibo and H. Ito, J. Theor. Biol. {\bf 378} 89 (2015).

% \bibitem{Takahashi2001}
% D. Takahashi, A. Shida, and M. Usami, 
% J. Phys A: Mathematical and General, 
% {\bf 23} 10715 (2001).

% \bibitem{Prigogine}
% G. Nicolis and I. Prigogine, \emph{Self-organization in nonequilibrium systems} (Wiley, New York, 1977).

% \bibitem{Guckenheimer}
% J. Guckenheimer and P. Holmes, \emph{Nonlinear Oscillations, Dynamical Systems, and Bifurcations of Vector Fields} (Springer, New York, 1983).

% \bibitem{Strogatz}
% Steven. H. Strogatz, \emph{Nonlinear Dynamics and Chaos} (Westview Press, U.S. 1994).

% \bibitem{Nicolis}
% G. Nicolis, \emph{Introduction to Nonlinear Science} (Cambridge Univ. Press 1995).

% \bibitem{Murray} 	
% J. D. Murray, \emph{Mathematical Biology} (Springer-Verlag, Berlin Heidelberg, 2002).

% \bibitem{Robinson}
% C. Robinson, \emph{Dynamical systems -Stability, Symbolic Dynamics, and Chaos-, 2ed edition} (CRC Press, Florida 1999).

% \bibitem{Kuznetsov}
% Yuri A. Kuznetsov, \emph{Elements of Applied Bifurcation Theory} (Springer-Verlag, New York, 2010).

% \bibitem{Grammaticos1997}
% B. Grammaticos, Y. Ohta, A. Ramani, D. Takahashi, and K. M. Tamizhmani, Phys. Lett. A {\bf 226}, 53 (1997).

% \bibitem{Takahashi1990}
% D. Takahashi and J. Satsuma, J. Phys. Soc. Jpn. {\bf 59} 3514 (1990).

% \bibitem{Matsukibara1997}
% J. Matsukibara, J. Satsuma, D. Takahashi, T. Tokihiro, and M. Torii, Phys. Lett. A {\bf 225} 287 (1997).

% \bibitem{Nagai1999}
% A. Nagai, D. Takahashi, and T. Tokihiro, Phys. Lett. A {\bf 255} 265 (1999).




% \bibitem{Nagatani1998}
% T. Nagatani, Phys. Rev. E {\bf 58} 700 (1998).



% \bibitem{Ohmori2014}
% S. Ohmori and Y. Yamazaki, Prog. Theor. Exp. Phys. 08A01 (2014).

% \bibitem{Matsuya2015}
% K. Matsuya and M. Murata, Discrete Contin. Dyn. Syst. B {\bf 20} 173 (2015).

% \bibitem{Murata2015}
% M. Murata, J. Phys. A Math, Theor. {\bf 48} 255202 (2015).


% \bibitem{Ohmori2016}
% S. Ohmori and Y. Yamazaki, J. Phys. Soc. Jpn. {\bf 85} 045001 (2016).

% \bibitem{Selkov1968}
% E. E. Sel$^{\prime}$kov, Eur. J. Biochem. {\bf 4} 79 (1968).

% \bibitem{Lengyel1990}
% I. Lengyel, G. Rabai, and I. R. Epstein, J. Am. Chem. Soc, {\bf 112} 9104 (1990).

% \bibitem{Lengyel1991}
% I. Lengyel and I. R. Epstein, Science, {\bf 251} 650 (1991).



%\bibitem{Baccelli}
%F. Baccelli, G. Cohen, G. J. Olsder, and J. P. Quadrat, \emph{Synchronization and Linearization} (Wiley, New York, 1992).

%\bibitem{Devaney}
%R. Devaney, \emph{An Introduction to Chaotic Dynamical Systems, 2ed edition} (Addison-Wesley 1989).



%\bibitem{FNeq}
%C. Rocsoreanu, A. Georgescu, and N. Giurgiteanu, \emph{The FitzHugh-Nagumo Model} (Springer Science+Business Media, Dordrecht, 2000). 

%\bibitem{Sasaki2018}
%M. Sasaki, S. Nishioka, F. Hongo, and M. Murata, Reports of RIAM Symposium, No.29AO-S7, 81 (2018). 

%\bibitem{Melo}
%W. Melo and R. G. Riezman, \emph{One-Dimensional Dynamics} (Springer-Verlag, New York, 2012).






%
\end{thebibliography}
\end{document}